\documentclass[aps,prd,citeautoscript,twocolumn,groupedaddress,amsfonts,amssymb,amsmath,showpacs,showkeys,nofootinbib]{revtex4-1}

\usepackage{eucal}

\usepackage{graphicx}

\usepackage{natbib}

\newcommand{\apss}{Astrophysics and Space Science}

\newcommand{\mnras}{MNRAS}

\newcommand{\lmatt}{\mathcal{L}_\text{matt}}

\begin{document}



\title{MOND as the weak field limit of an extended metric theory of gravity
with a matter-curvature coupling.}

\author{E. Barrientos}
\email[Email address: ]{ebarrientos@astro.unam.mx}
\author{S. Mendoza}
\email[Email address: ]{sergio@astro.unam.mx}
\affiliation{Instituto de Astronom\'{\i}a, Universidad Nacional
                 Aut\'onoma de M\'exico, AP 70-264, Ciudad de M\'exico, 04510,
	         M\'exico \\
            }

\date{\today}

\begin{abstract}
  In this article we construct an extended relativistic \( f(R) \) theory
of gravity with matter-curvature couplings \( F(R,\lmatt) \) for which its
weak field limit of approximation recovers the simplest version of MOND.
We do this by (a) performing an order of magnitude approach and (b) by
perturbing the resulting field equations of the theory to the weakest field
limit of approximation.  We also compute the geodesic equation of the
resulting theory and show that it has an extra force, a fact that commonly
appears in general matter-curvature couplings. 
\textbf{This arxiv version includes a section at the end with an Erratum.}
\end{abstract}


\maketitle

\section{Introduction}
\label{Introduccion}

  The non-baryonic dark matter problem constitutes one of the most
important unsolved problems in current research \citep[cf.][]{dm01,dm02}.
Despite the huge research and its generally accepted success, the
dark matter particle has never been detected.  The gravitational
anomaly that gives rise to the dark matter and/or energy hypothesis
can also be understood as a modification of gravity at certain scales
\citep[cf.][]{mendoza15} as it was first discussed by the pioneer research
of \citet{Milgrom1, Milgrom2}, with a MOdified Newtonian Dynamics (MOND)
approach.  A first coherent attempt to find a relativistic version was
carried out by \citet{Bekenstein} with a TEnsor Scalar Vector (TEVES)
theory, this idea has been widely explored \citep{Bek2, Sanders-t-v-s,
scalar-tensor-sanders,Zlosnik2, SkortisTeVeS},  but due to the extreme
complexity of the theory and some clear failures, research has continued
into finding a relativistic theory of gravity which yields MOND in its
non-relativistic, weakest field limit regime.

  \citet{bernal11} showed that MOND acceleration can be accounted by a
relativistic  \(f(\chi)=\chi^{3/2}\) metric theory of
gravity described by the action:

\begin{equation}
     S  =  \frac{ c^3 }{ 16 \pi G L_M^2 } \int{ f(\chi) \sqrt{-g}
       \, \mathrm{d}^4x}  + \frac{1}{c} \int{ \mathcal{L}_\text{matt} \sqrt{-g}
       \, \mathrm{d}^4x},
\label{fchi}
\end{equation}

\noindent where \( \chi := L^2 R \), \( R \) is the Ricci scalar, \(
L \propto r_\text{g}^{1/2} l^{1/2} \), with \(r_\text{g}:= GM/c^2 \)
the gravitational radius,  \( l:= \left( GM / a_0 \right)^{1/2} \) the
``mass-length'' scale of the system and \( \mathcal{L}_\text{matt}
\) is the standard matter Lagrangian, related to the energy-momentum tensor
\( T_{\alpha\beta} \) by: 

\begin{equation}
  T_{\alpha\beta} \sqrt{-g} \, \delta g^{\alpha\beta} = -  2 \delta \left(
    \sqrt{-g} \, \mathcal{L}_\text{matt} \right).
\label{energymomentum}
\end{equation}

\noindent The constant \( a_0 \approx 1.2 \times 10^{-10} \textrm{m}\,
\textrm{s}^{-2} \) is Milgrom's acceleration constant. This proposal
is coherent with the results of gravitational lensing in individual,
groups and clusters of galaxies \citep{mendoza13} and at the same second 
perturbation order is coherent with a Parametrised Post-Newtonian (PPN)
description where the parameter \( \gamma = 1 \) \citep{mendozaolmo}.
Another extension of gravity was performed by \citet{Barr1}, who analysed
the action \eqref{fchi} but now using the Palatini approach, obtaining
the same functional action $f(\chi)=\chi^{3/2}$ in order to recover the
MONDian acceleration, with a  mass dependence on the coupling
length \( L \).

  The problem with action~\eqref{fchi} is that it can only be applied
in regions sufficiently far from the sources that produce the gravitational
field, in order to approximate the system as a point mass source.  There is
however a cosmological attempt by \citet{carranza13} in which the mass 
\( M \) was thought of as the causal mass for a particular observer in 
the cosmic flow, yielding a good description of an accelerated expansion of
the universe without the introduction of dark matter and/or energy.  

  Another recent exploration was carried out by \citet{Barr2} who
showed that the mass dependence in the coupling length $L$ can be
avoided introducing derivatives of the matter Lagrangian in the action
$f(\chi)$. In such proposal the coupling constant depends exclusively on
the fundamental constants $c$, $a_0$ and $G$, but the price to pay is in
the complexity of the field equations and the theoretical inconvenients
that the introduction of the derivatives of the matter  Lagrangian produce.

  In this article we use an extension of a metric \( f(R) \)  theory of 
gravity with matter-curvature couplings \( F(R,\lmatt)
\) following the approach by \citep{frlm, Lobo, Harko1, Harko3, Harko4} 
and show that with this generalised action a relativistic theory of MOND 
can be constructed. The article is presented in the following manner.  In
Section~\ref{approach} an order of magnitude calculation is performed to
show that a specific \( F(R,\lmatt) \) can reproduce MOND in its simplest
form.  Section~\ref{mondlimit} shows an exact solution for a point-mass
source reproducing these results.  In Section~\ref{dimensional} we use
correct dimensional arguments to generalise an action for a \( F(R,\lmatt)
\) and show that with this it is possible to recover either  MOND or
Newton's gravity at the weakest field limit of the theory.  Finally in
Section~\ref{discussion} we discuss the results of the article and
present our conclusions.

\section{$F(R,L_\text{matt})$ approach}
\label{approach}

  The lesson to learn from action~\eqref{fchi} is that the matter
Lagrangian \( \lmatt \) needs to be inserted inside the gravitational
action (see e.g. \citet{mendoza15}). The idea of a non-minimal coupling 
between the matter and the curvature has been already raised 
\citep{Goenner, Nojiri, Allemandi, Bertolami}.
 To do so, we can use an extension of  \(
f(R) \) of gravity introducing a \( F(R,\lmatt) \) described by \citet{frlm}:

\begin{equation}
  S = \int{F(R,\lmatt) \sqrt{-g} \, \mathrm{d}^4 x }, 
\label{action}
\end{equation}

\noindent with the following field equations:

\begin{equation}
  \begin{split}
  F_R R_{\alpha\beta} +& \left( g_{\alpha\beta} \nabla^\mu \nabla_\mu -
  \nabla_\alpha \nabla_\beta \right) F_R \\
    & - \frac{ 1 }{ 2 } \left( F -
  F_{\lmatt} \right)g_{\alpha\beta} = \frac{ 1 }{ 2 } F_{\lmatt}
  T_{\alpha\beta},
\end{split}
\label{fieldeqs}
\end{equation}

\noindent where \( F_R := \partial F / \partial R  \) and \( F_{\lmatt} :=
\partial F / \partial \lmatt \).  Note that (a) \( F(R,\lmatt)= c^3 R /
16\pi G + \lmatt / c \) yields standard general relativity, (b) \( F(R,\lmatt)
= f(R)/2 + \lmatt / c \) is standard metric \( f(R) \) gravity and (c):

\begin{equation}
  F(R,\lmatt) = \frac{c^3}{ 16 \pi G } \frac{ f(\chi) }{ L^2 } +
  \frac{1}{c} \lmatt,
\label{mond}
\end{equation}

\noindent is a correct generalisation of~\eqref{fchi} in which the unknown
length function \( L = L(\lmatt) \) is to be found; and together with the
unknown function\( f(\chi) \) must yield a correct MOND behaviour in the
limit of low acceleration scales \( a \lessapprox a_0 \).

\section{MONDian limit}
\label{mondlimit}

  Let us now show that with the assumptions made in section~\ref{approach}
it is possible to obtain the basic MOND relation based on the
Tully-Fisher law.  To do so, let us substitute equation~\eqref{mond}
into the field equations~\eqref{fieldeqs} and take the trace of the
resulting relation to yield:

\begin{equation}
  f_R(\chi) R +- 2f(\chi) + 3 L^2 \nabla^\alpha\nabla_\alpha \left(
    \frac{ f_R(\chi) }{ L^2 }  \right) = \frac{ 8 \pi G L^2 }{ c^4 } 
    T^\alpha_{{\ }\alpha}.
\label{trace}
\end{equation}

\noindent In order to find the correct MONDian limit equation, we follow the
procedure by \citet{bernal11} and so, let 

\begin{equation}
  f(\chi) = \chi^b, \qquad \text{and} \qquad \lmatt = \rho c^2, 
\label{nolabel}
\end{equation}

\noindent where we have assumed a point mass source generating the
gravitational field, and so \( \lmatt \) has a dust-like form.  
To order of magnitude, i.e. when \( R
\sim r_\text{curv}^{-2} \) -where \(r_\text{curv} \) is the radius of
curvature of space- and \( \nabla \sim 1 / r \), it follows 
that the first two terms on
the left-hand side of equation~\eqref{trace} are smaller than the third
when \( r/r_\text{curv} \rightarrow 0 \), i.e. when the equivalent
acceleration \( a \) is expected to be \( \lesssim a_0 \).

  Thus, the trace of the field equations that can be adapted to a
MONDian regime of low acceleration scales is given by:

\begin{equation}
  3 L^2 \nabla^\alpha\nabla_\alpha \left(
    \frac{ f_R(\chi) }{ L^2 }  \right) = \frac{ 8 \pi G L^2 }{ c^4 } 
    T^\alpha_{{\ }\alpha}.
\label{tracegood}
\end{equation}

  A weak-field limit coherent with bending of light in individual, groups
and clusters of galaxies is obtained if the second perturbation order 
metric is given by \citep{mendozaolmo}:

\begin{equation}
  \mathrm{d}s^2 = \left( 1 + \frac{ 2 \phi }{ c^2 } \right) 
  \, c^2 \mathrm{d}t^2 - \left( 1 -
  \frac{ 2 \phi }{ c^2 } \right) \, \mathrm{d}\boldsymbol{x}^2,
\end{equation}

\noindent for a gravitational scalar potential \( \phi \) and an isotropic
space-time with a PPN parameter \( \gamma \approx 1 \) according to
observations of such MONDian systems \citep{mendoza13}.
With this, the Ricci scalar takes the form: \( R \approx - ( 2 / c^2 )
\nabla^2 \phi \), which at order of magnitude yields: \( R \sim a /
rc^2 \), for an acceleration \( a = | \nabla \phi | \).

  Thus, to order of magnitude, equation~\eqref{tracegood} yields:

\begin{equation}
  a \sim G^{1/(b-1)}\, \rho^{1/(b-1)}\,  r^{(b+1)/(b-1)}\, 
    c^{(2b-4)/(b-1)}L^{-2},
\label{ordermag}
\end{equation}

\noindent and so, in order to obtain MOND standard equation: \( a =
\sqrt{G\, a_0\, M} / r \sim  \sqrt{G \, a_0 \, \rho \, r} \), then \( b=-3 \)
together with \( L \propto \left( G \rho \right)^{-3/8} c^{5/4} a_0^{1/4}
\), which yields:

\begin{equation}
  F(R,\lmatt) \propto R^{-3} \lmatt^{3}.
\end{equation}

\section{A dimensionally correct general action}
\label{dimensional}

  Let us now consider an action motivated by equation~\eqref{fchi} 
with the following form:

\begin{equation}
     S  =  \frac{ c^3 }{ 16 \pi G \alpha } \sqrt{-g} \int{ f(\chi , \xi)
       \, \mathrm{d}^4x} + \frac{1}{c} \int{\sqrt{-g} \mathcal{L}_\text{matt} 
       \, \mathrm{d}^4x},
\label{fchi2}
\end{equation}

\noindent where $\chi$ and $\xi$ are dimensionless quantities given by:

\begin{equation}
	\xi := \frac{\mathcal{L}_\text{matt}}{\lambda}, \qquad 
	  \text{and} \qquad  \chi := \alpha R,
	\label{adimen}
\end{equation}

\noindent with $\alpha$ and $\lambda$ unknown ``coupling'' constants
with dimensions of square length and energy
density respectively.

The null variations with respect to the metric yields the following field
equations:

\begin{equation}
	\begin{split}
	&\alpha f_\chi R_{\mu\nu}-\frac{1}{2}g_{\mu\nu}(f-\xi f_\xi)\\
	&=\left(\frac{8\pi G \alpha}{c^4}+\frac{f_\xi}{2\lambda}\right)T_{\mu\nu}-\alpha(g_{\mu\nu}\Delta-\nabla_\mu \nabla_\nu) f_\chi.
	\end{split}
	\label{fieldadim}
\end{equation}

\noindent with the standard definition of the energy-momentum tensor:

\begin{equation}
	T_{\mu\nu}=g_{\mu\nu}\mathcal{L}_\text{matt}-2 \frac{\partial
	\mathcal{L}_\text{matt}}{\partial g^{\mu\nu}},
	\label{Tmunu}
\end{equation}

\noindent in full agreement with equation~\eqref{energymomentum}.

  The trace of equation~\eqref{fieldadim} is given by:

\begin{equation}
	\chi f_\chi-2(f-\xi f_\xi)+3\alpha\Delta f_\chi=\left(\frac{8\pi G \alpha}{c^4}+\frac{f_\xi}{2\lambda}\right)T.
	\label{trazaadim}
\end{equation}

  Since $c$, $G$ and $a_0$ are independent fundamental constants,
Buckingham's \( \Pi \) theorem of dimensional analysis implies that:

\begin{equation}
  \alpha=\kappa\frac{c^4}{{a_0}^2} \qquad \text{and} \qquad
\lambda=\kappa'\frac{{a_0}^2}{G},
\label{Buck}
\end{equation}

\noindent with \( \kappa \) and \( \kappa' \) pure dimensionless
proportionality constants.

 Following the previous approach, we can assume that:

\begin{equation}
	f(\chi , \xi)= \chi^\gamma \xi^\beta.
	\label{potencias}
\end{equation}

\noindent  For the case of dust, the perturbation orders in the terms of 
the field equation are the following:

\begin{equation}
	\begin{split}
	&\overbrace{\alpha f_\chi R_{\mu\nu}-\frac{1}{2}g_{\mu\nu}(f-\xi f_\xi)}^{{\cal{O}}(-2(\gamma+\beta))}
		+\overbrace{\alpha(g_{\mu\nu}\Delta-\nabla_\mu \nabla_\nu) f_\chi}^{{\cal{O}}(-2(\gamma+\beta+1))}\\
	&=\underbrace{\frac{8\pi G \alpha}{c^4}T_{\mu\nu}}_{{\cal{O}}(2)}+\underbrace{\frac{f_\xi}{2\lambda}T_{\mu\nu}}_{{\cal{O}}(2(\gamma+\beta))}.
	\end{split}
	\label{orders}
\end{equation}

\subsection{Poisson-like equation for MOND}

  The lowest perturbation order of the previous equation is \( 2 \) and so,
the choice  \( \gamma = - \beta \) yields:

\begin{equation}
  \left( g_{\mu\nu}\Delta-\nabla_\mu \nabla_\nu \right) f_\chi=\frac{8\pi
    G}{c^4}T_{\mu\nu}.
\label{primerael}
\end{equation}

  Contracting equation~\eqref{primerael} with $g^{\mu\nu}$ gives:

\begin{equation}
	3\Delta f_\chi=\frac{8\pi G}{c^4}T,
	\label{traza1}
\end{equation}

\noindent which at the lowest perturbation order for dust takes 
the following expression:

\begin{equation}
	(-2\kappa)^{\gamma-1}\kappa'^\gamma \frac{{a_0}^2}{G^{\gamma+1}}\nabla^2\left(\left\{\nabla^2 \phi\right\}^{\gamma-1}\rho^{-\gamma}\right)
		=\frac{8\pi}{3}\rho.
	\label{poissonMOND}
\end{equation}

\noindent To order of magnitude, this last equation implies that:

\begin{equation}
	a\approx M^{(1+\gamma)/(\gamma-1)}r^{-2(1+\gamma)/(\gamma-1)},
	\label{magnitude}
\end{equation}

\noindent and so, in order to recover a MONDian expression for the 
acceleration, the following value of $\gamma$ is found:

\begin{equation}
	\gamma=-3.
	\label{gamma}
\end{equation}

\noindent With this value, the Poisson-like equation~\eqref{poissonMOND} is:

\begin{equation}
	\frac{3}{8\pi}\frac{(a_0G)^2}{(2\kappa)^4\kappa'^3}\nabla^2\left(\left\{\nabla^2 \phi\right\}^{-4}\rho^3\right)=\rho.
	\label{poissonMOND2}
\end{equation}

  An analytic solution to the previous equation for the case of a
point-mass source is given in the appendix \ref{Apen}. 

  Note that equation~\eqref{poissonMOND2} represents a non-linear
generalisation of the standard Poisson equation \( \nabla^2 \phi \propto
\rho \).  A family of these non-linear generalisations was discussed  by
\cite{Milgrom97}, with Poisson-like equations
of the form \( \nabla \cdot \left( \mu\left( | \nabla \phi | \right)  
\nabla \phi \right) \propto \rho \) satisfying conformal invariance in all
cases studied.  Equation~\eqref{poissonMOND2} does not fall into that
category and as such, it differs from the standard AQUAL proposal
\citep{bekenstein84}.  This is due to the fact that the nonlinearity of
equation~\eqref{poissonMOND2} does not only apply to the scalar potential
\( \phi \) but also to the mass density \( \rho \), since this last one
appears inside the Laplacian operator on the left-hand side of
relation~\eqref{poissonMOND2}.

\subsection{Poisson's equation for Newtonian gravity.}

  Another possible choice for equation~\eqref{orders} is \( \gamma + \beta = 1
\) which yields:

\begin{equation}
	\alpha f_\chi R_{\mu\nu}-\frac{1}{2}g_{\mu\nu}(f-\xi f_\xi)
		=\left(\frac{8\pi G
		\alpha}{c^4}+\frac{f_\xi}{2\lambda}\right)T_{\mu\nu}.
	\label{segundael}
\end{equation}

\noindent This lowest perturbation order choice means that:

\begin{equation}
	(g_{\mu\nu}\Delta-\nabla_\mu \nabla_\nu) f_\chi=0.
	\label{consec}
\end{equation}

  Taking the trace of equation~\eqref{segundael} for dust, a relation
between the Ricci scalar and the matter density is obtained:

\begin{equation}
   R=\left(-\frac{16\pi}{\gamma+1}(\kappa\kappa')^{1-\gamma}
     \right)^{1/\gamma}\frac{G}{c^2}\rho.
\label{PoissonN}
\end{equation}

\noindent At the lowest perturbation order, when \( R = - ( 2 / c^2 )
\nabla^2 \phi \), this previous equation can be constructed -with the
appropriate coupling constants- to yield Newtonian gravity (Poisson's
equation) for any value of \( \gamma \neq -1 \).

\section{Discussion}
\label{discussion}

  In this article we have shown that it is possible to  show, exactly and
by an order of magnitude approach,  that a \( F(R,\lmatt) \) theory of 
gravity described by:

\begin{equation}
 f(\chi, \xi\,) = \chi^{-3}\xi^3, \qquad \chi := \alpha R, \quad
   \xi:=\mathcal{L}_\text{matt}/\lambda,
\end{equation}

\noindent is a good candidate for a full relativistic extension of
MOND, in regions where the acceleration of test particles
\( \lesssim a_0 \).  In the weak-field limit of approximation it
converges to standard MOND for a point mass source \( M \), with 
\( \rho = M \delta(\boldsymbol{r}) \) and 
\( \lmatt= \rho c^2 \). It is our intention to explore
this interpretation with applications to lensing and dynamics of
individual, groups and clusters of galaxies as well as with cosmology.
The advantage of this approach is that it is a full metric formalism and
does not involve interpretations of gravity using Palatini formalism or
torsion as we have previously explored \citep{barrientos16,barrientos17}.
Furthermore, it is a correct generalisation to the first attempts made
by \citet{bernal11}.

  At first sight, the action given by the Lagrangian density:
$R^{-3}\mathcal{L}_\text{matt}^3$ from which we have proved the MONDian
behaviour is obtained, seems to diverge in the Minkowskian regime, namely
when $R\rightarrow 0$. In order to show that this is not so, we proceed
in the following way.  Using relations \eqref{Buck}, \eqref{potencias},
\eqref{gamma}, and the fact that \( \gamma = -\beta \), 
expression~\eqref{traza1} turns into:

\begin{equation}
	-\frac{9}{8\pi k^4 k'^3}\left(\frac{a_0 G}{c^6}\right)^2 \Delta(R^{-4}\mathcal{L}_\text{matt}^3)=T,
	\label{diverg}
\end{equation}

\noindent which in the weak-field limit for a point-mass source is:

\begin{equation}
	-\frac{9}{8\pi k^4 k'^3}\left(\frac{a_0 G}{c^5}\right)^2
	\nabla^2(R^{-4}\mathcal{L}_\text{matt}^3)=M\delta(\boldsymbol{r}).
	\label{diverg2}
\end{equation}

\noindent Using the well known result: 

\begin{equation}
	\nabla^2\left(\frac{1}{ \boldsymbol{r} }\right)=-4\pi
	\delta(\boldsymbol{r}),
	\label{laplacian}
\end{equation}

\noindent the following relation is satisfied:

\begin{equation}
	R^{-4}\mathcal{L}_\text{matt}^3=\frac{2\pi k^4 k'^3}{9}\left(\frac{c^5}{a_0 G}\right)^2\frac{M}{r}.
	\label{diver3}
\end{equation}

Therefore, in the weak field limit,  this proposal has the following relation: 
$\mathcal{L}_\text{matt}^3\propto R^4/r$. This implies that the Lagrangian 
density for the action that we are interested in converges to
$R^{-3}\mathcal{L}_\text{matt}^3\propto R/r \rightarrow 0 $ as \(
r \) increases.

Finally, we discuss the geodesic equation of the theory. Following a
similar procedure as the one shown in \citep{Koivisto,frlm}, the geodesic
equation is given by:

\begin{equation}
    \frac{\mathrm{d}
      x^\mu}{\mathrm{d}s^2}+\Gamma^\mu\,_{\nu\alpha}
      \frac{\mathrm{d}x^\nu}{\mathrm{d}
      s}\frac{\mathrm{d}x^\alpha}{\mathrm{d}s}=f^\mu,
\label{geodesic}  
\end{equation}

\noindent where

\begin{equation}
    f_\mu= \left( g_{\mu\nu}-u_\mu u_\nu\right)\nabla^\mu \text{ln}\left[\left(16\pi\kappa\kappa'+f_\xi\right)
        \frac{\mathrm{d}{\mathcal{L}_\text{matt}}}{\mathrm{d}\rho}\right].
    \label{fuerza}
\end{equation}

As expected, the usual relation $u_\mu f^\mu=0$ is obtained. This means
that the extra force is perpendicular to the four-velocity.  For dust,
the extra-force takes the following form:

\begin{equation}
    f_\nu= \left( g_{\mu\nu}-u_\mu u_\nu\right)\nabla^\mu \text{ln}\left[16\pi\kappa\kappa'+f_\xi\right].
    \label{fuerza2}
\end{equation}

\noindent This type of extra force has been studied and interpreted in
the literature \cite[cf.][]{harko14} and in a very different context to
the one discussed in this article to yield MOND-like accelerations by
\cite{Vagnozzi}.  Investigations into its nature and its astrophysical
consequences requires further research.

\section*{Acknowledgements}
This work was supported
by DGAPA-UNAM (IN112616) and CONACyT (CB-2014-01 No.~240512) 
grants. EB and SM acknowledge economic support from CONACyT 
(517586 and 26344).

\appendix
\section{Poisson-like equation}
\label{Apen}

 Let us begin by rewriting equation \eqref{poissonMOND2} as:

\begin{equation}
	K\nabla^2\left(\left\{\nabla^2 \phi\right\}^{-4}\rho^3\right)=\rho,
\label{Poisson}
\end{equation}
\noindent where for simplicity we have defined:

\begin{equation}
	K := \frac{3}{8\pi}\frac{(a_0G)^2}{(2\kappa)^4\kappa'^3}.
\label{constante}
\end{equation}

  The matter density for a point-mass source is given by:

\begin{equation}
	\rho=\frac{M}{4\pi r^2}\delta(r),
\label{density}
\end{equation}
\noindent and since the Laplacian for a spherically symmetric problem is:

\begin{equation}
	\nabla^2 \psi=\frac{1}{r^2}\frac{ \mathrm{d} }{ \mathrm{d} r }
	  \left(r^2\frac{d\psi}{dr}\right),
\label{laplacian2}
\end{equation}

\noindent then, equation~\eqref{Poisson} turns into:

\begin{equation}
  4\pi K \frac{ \mathrm{d} }{\mathrm{d} r}\left( r^2 \frac{\mathrm{d}
  }{ \mathrm{d} r} \left(\left\{\nabla^2 \phi\right\}^{-4}
  \rho^3\right)\right)=M\delta(r).
\label{Proc1}
\end{equation}

\noindent Integration of the previous equation yields:

\begin{equation}
	4\pi K \frac{\mathrm{d}}{\mathrm{d}r}\left(\left\{\nabla^2
	\phi\right\}^{-4}\rho^3\right)=\frac{M}{r^2},
\label{Proc2}
\end{equation}

\noindent which after another integration gives:

\begin{equation}
	4\pi K\left\{\nabla^2 \phi\right\}^{-4}\rho^3=-\frac{M}{r}.
	\label{Proc2.1}
\end{equation}

Using again eqs. \eqref{density} and \eqref{laplacian2} and after a some
algebraic steps, we obtain:

\begin{equation}
  (-K)^{1/4}\left(\frac{ M }{ 4 \pi} \right)^{1/2} \left( \frac{r^3}{
    \delta(r)} \right)^{1/4} \delta(r) = \frac{\mathrm{d}}{\mathrm{d}
    r}\left(r^2\frac{ \mathrm{d} \phi}{\mathrm{d} r}\right),
\label{Proc3}
\end{equation}

\noindent which after another integration is written as:

\begin{equation}
  \left.(-K)^{1/4} \left( \frac{M}{4\pi }\right)^{1/2} \left( \frac{r^3
    }{ \delta(r) } \right)^{1/4} \right|_0 = r^2 \frac{\mathrm{d}
    \phi}{\mathrm{d} r}.
\label{Proc4}
\end{equation}

  Using the fact that the acceleration $ a = | \boldsymbol{a} | = | \nabla \phi
| $ and the Dirac's delta function is given by:

\begin{equation}
	\delta(r=0) = \lim_{r\rightarrow 0}\frac{1}{2\pi r},
	\label{delta0}
\end{equation}
\noindent then the relation for the accelerations is given by:

\begin{equation}
	\left(-K\frac{M^2}{2^3\pi}\right)^{1/4}\frac{1}{r}= a.
	\label{Proc5}
\end{equation}

Substitution of the value of $K$ given in equation~\eqref{constante},
yields to:

\begin{equation}
  \left( -\frac{ 3 }{ 4^5 \kappa'^3\pi^2 }\right)^{1/4} \frac{1}{\kappa}
    \frac{(a_0 GM )^{1/2}}{ r } = a.
\label{aceleracion}
\end{equation}

\noindent Thus, the choice \( \kappa'^3  = - 3 / 4^5 \pi^2 \kappa'^4 \) yields a MONDian 
acceleration $a = \sqrt{GMa_0}/r$.

\clearpage

\onecolumngrid
\begin{center}
\textbf{Erratum to: MOND as the weak field limit of an extended metric theory of gravity
with a matter-curvature coupling.} \\
\bigskip
In article \cite{acoplamiento}, the right-hand side of equation \eqref{poissonMOND} 
had a missing \(\gamma \) factor.  The error propagates further and the coupling constants
\( \kappa \) and \( \kappa' \) have different values with this corrections.
Furthermore, there are two small errors in equations \eqref{fieldeqs} and \eqref{trace} which do not 
have any propagation in the results obtained in the article. In this erratum we show the correct expressions.
\end{center}

\twocolumngrid

The field equations \eqref{fieldeqs} for the proposal have a term $\lmatt$ missing. The correct 
equations are given by: 

\begin{equation}
  \begin{split}
  F_R R_{\alpha\beta} +& \left( g_{\alpha\beta} \nabla^\mu \nabla_\mu -
  \nabla_\alpha \nabla_\beta \right) F_R \\
    & - \frac{ 1 }{ 2 } \left( F -
  \lmatt F_{\lmatt} \right)g_{\alpha\beta} = \frac{ 1 }{ 2 } F_{\lmatt}
  T_{\alpha\beta}.
\end{split}
\label{fieldeqs_err}
\end{equation}

On the other hand, the trace of the latter relation is:

\begin{equation}
  f_R(\chi) R - 2f(\chi) + 3 L^2 \nabla^\alpha\nabla_\alpha \left(
    \frac{ f_R(\chi) }{ L^2 }  \right) = \frac{ 8 \pi G L^2 }{ c^4 } 
    T^\alpha_{{\ }\alpha}.
\label{trace_err}
\end{equation}

\noindent where an extra $+$ sign has been removed in eq. \eqref{trace}.

The main error made in \cite{acoplamiento} was the missing of $\gamma$ in equation \eqref{poissonMOND}. 
This term has its origin in $f_\chi=\gamma \chi^{\gamma-1}\xi^{-\gamma}$. Taking this into account, 
the field equations for the lowest perturbation in the case of dust is given by: 

\begin{equation}
	(-2\kappa)^{\gamma-1}\kappa'^\gamma \gamma \frac{{a_0}^2}{G^{\gamma+1}}\nabla^2
	\left(\left\{\nabla^2 \phi\right\}^{\gamma-1}\rho^{-\gamma}\right)=\frac{8\pi}{3}\rho.
	\label{poissonMOND_err}
\end{equation}

The lack of this term does not have repercussion in the obtained value for $\gamma$ but it does 
in the value of the coupling constants $\kappa$ and $\kappa'$. Equation \eqref{constante} in the 
appendix must be replaced by: 

\begin{equation}
	K :=- \frac{9}{8\pi}\frac{(a_0G)^2}{(2\kappa)^4\kappa'^3}.
\label{constante_err}
\end{equation}

The mistake propagates to equation \eqref{aceleracion}, which when corrected turns into: 

\begin{equation}
  \left( \frac{ 9 }{ 4^5 \kappa'^3\pi^2 }\right)^{1/4} \frac{1}{\kappa}
    \frac{(a_0 GM )^{1/2}}{ r } = a.
\label{aceleracion_err}
\end{equation}

Therefore, the relation for the coupling constants which allows to recover the MONDian 
acclerations is given by: $\kappa^4\kappa'^3  = 9 / 4^5 \pi^2$.

\bibliographystyle{aipnum4-1}
\bibliography{barrientos_mendoza}

\end{document}